\documentstyle[twoside,epsfig,psfig]{article} 
 
\input ibvs2.sty 
 
\begin{document} 
\IBVShead{6162}{8 Mar 2015} 
 
\IBVStitle{Photometric evolution of the 2016 outburst}
\vskip -0.3cm
\IBVStitle{of recurrent Nova LMC 1968: the first three weeks}
\IBVStitle{} 
  
\IBVSauth{U. Munari$^1$, F. M. Walter$^2$, F.-J. Hambsch$^3$, A. Frigo$^3$} 

\IBVSinst{INAF Osservatorio Astronomico di Padova, Sede di Asiago, I-36032 Asiago (VI), Italy} 
\IBVSinst{Dept. of Physics and Astronomy, Stony Brook University, Stony Brook, NY 11794-3800, USA} 
\IBVSinst{ANS Collaboration, c/o Astronomical Observatory, 36012 Asiago (VI), Italy}

\IBVStyp{Nova} 
\IBVSkey{photometry} 
\IBVSabs{Optical (BVRI) photometry of the first three weeks of the 2016
outburst of the recurrent Nova LMC 1968 is presented and discussed.  The
2016 I-band light-curve is an exact replica, even in the most minute
details, of that for the 2010 eruption.  The maximum is inferred to have
occurred on 2016 Jan 21.2 at I=11.5 mag, corresponding to an absolute
magnitude $M_I$=-7.15.  A $\sim$1 day long plateau is present in all bands
about six days past optical maximum, simultaneous with the emergence of
super-soft X-ray emission in Swift observations, signalling the widespread
ionization of the ejecta.  The nova entered a much longer plateau about 9
days past maximum, governed by the brightness of the white dwarf, now
directly visible and still nuclearly burning on its surface.  A recurrence
period of $\sim$955 days would fit both the OGLE inter-season gaps and the
observed intervals between previous outburts.}

\begintext 

When Nova LMC 1968 (= LMCN 1968-12a = Nova Men 1968; Sievers 1970) erupted
again in 1990 (as Nova LMC 1990 N.2 = LMCN 1990-02a = LMC V1341; Liller
1990), it became the first recurrent nova known in LMC (Shore et al.  1991). 
Further outbursts of Nova LMC 1968 were observed in 2002 (by the ASAS-3
survey, Pojmanski 2002), and in 2010 (by the OGLE survey, Mroz et al. 
2014).  A fifth eruption, detected in OGLE real-time data, has just been
announced by Mroz and Udalski (2016).  In this paper we present and discuss
our extensive $B$,$V$,$R_{\rm C}$,$I_{\rm C}$ photometry of the first three
weeks of 2016 event, while further observations are in progress.

We have obtained optical photometry of the nova with the ($a$) SMARTS 1.3-m
telescope + ANDICAM from CTIO (Chile), which data are reduced against
nightly observations of all-sky standard stars (Walter et al.  2012), and
with ($b$) a 0.4-m robotic telescope operated by ANS Collaboration in
Atacama (Chile), which data are reduced against a local photometric sequence
extracted from the all-sky APASS survey (Henden et al.  2012, Munari et al. 
2014).  Two faint field stars are located within $\sim$2 arcsec of the nova,
a fact not appreciated in previous outbursts whose published photometry
refers to the combined (non-deconvolved) light of the nova and of these two
nearby field stars.  The combined magnitude of the two field stars is
$B$=18.79, $V$=18.26, $R_{\rm C}$=17.86, and $I_{\rm C}$=17.49.  In Figure~1
we present our non-deconvolved $B$,$V$,$R_{\rm C}$,$I_{\rm C}$ photometry of
the 2016 outburst (measured through a 11-arcsec aperture on both
telescopes), while deconvolved SMARTS photometry (PSF-fitting) is
plotted in Figure~2, and both sets of data are listed in Table~1 (available
electronically only).  No deconvolution is possible for SMARTS JD=2457414.65
observation because of the $\sim$3 arcsec seeing affecting it.  ANS
photometry is not deconvolved because of the focal length too short for a
meaningfull PSF-fitting.  In both figures, the continuous line is the
deconvolved OGLE $I_{\rm C}$-band photometry for the 2010 outburst, plotted
for reference.  A similar comparison with other outbursts and/or other
wavelengths is not possible because the published photometry is very scanty,
with only a few points being - at best - available per outburst.

The decline rates for the first week of the 2016 outburst are 0.57
mag~day$^{-1}$ in $R_{\rm C}$, 0.53 in $I_{\rm C}$, 0.50 in $B$, and 0.45 in
$V$.  The corresponding classical $t_2$ and $t_3$ decline times are 3.5 and
5.2 days in $R_{\rm C}$, 3.8 and 5.7 in $I_{\rm C}$, 4.0 and 5.9 in $B$, and
4.5 and 6.7 in $V$.  The 
\clearpage

\IBVSfig{16cm}{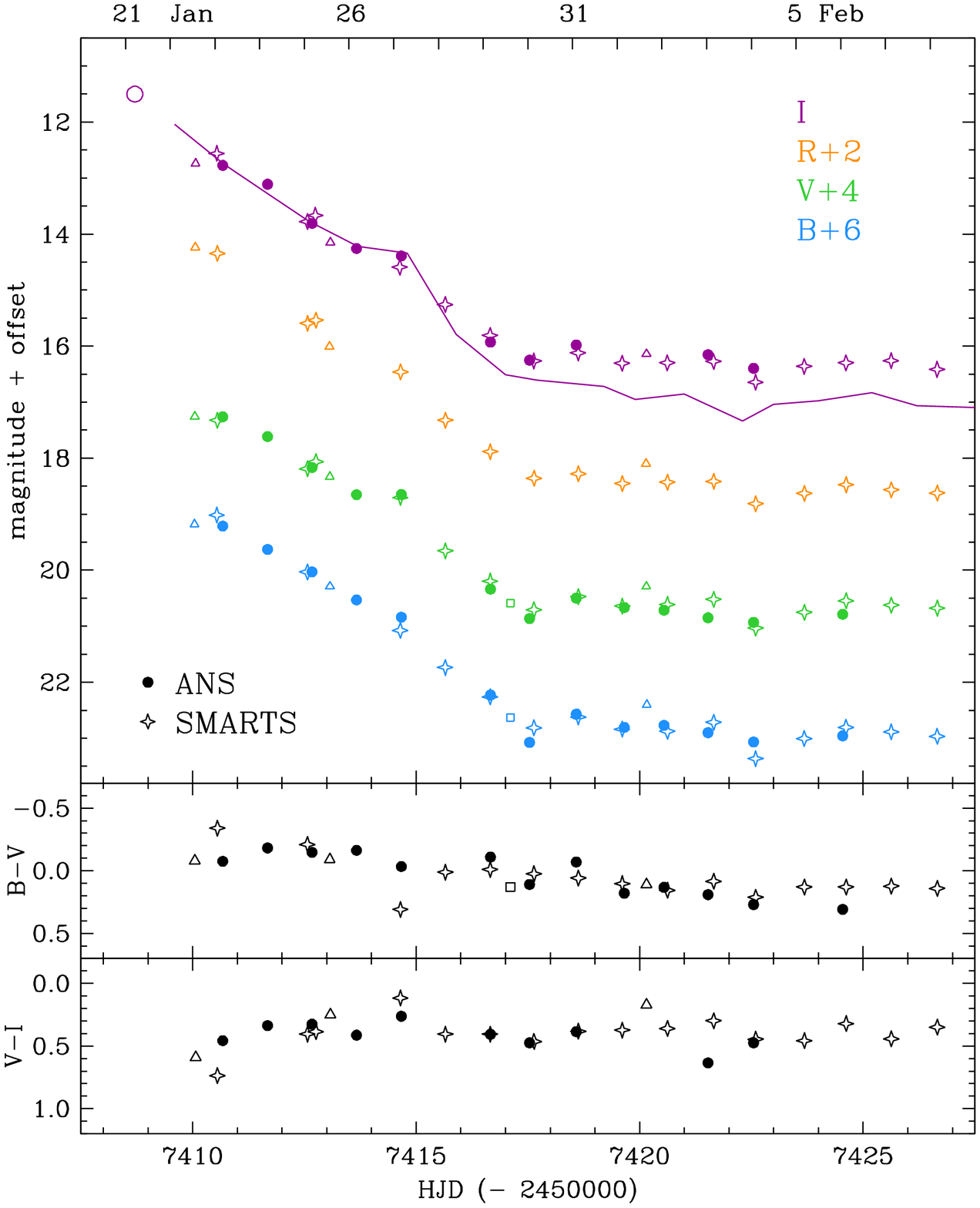}{Non-deconvolved $B$$V$$R_{\rm C}$$I_{\rm C}$
photometry of the 2016 outburst of Nova LMC 1968. The open circle is the
$I_{\rm C}$-band discovery observation by Mroz and Udalski (2016), and the 
line is the $I_{\rm C}$-band lightcurve of the 2010 outburst (from Mroz et
al. 2014). Triangles and squares are observations by Kaur et al. (2016)
and Darnley and Williams (2016), respectively.}

\noindent
2016 outburst is characterized by a striking similarity to that of 2010, at
least for the $I_{\rm C}$ light-curve, as illustrated in Figures~1 and 2. 
In addition to the identical decline rates, the $\sim$1 day {\em plateau}
the nova went through between 26 and 27 Jan 2016, is similarly present in
the 2010 light-curve, and coincides with the first appearance of super-soft
X-ray emission in 2016 Swift observations of the nova (Page et al.  2016). 
The detection of emerging super-soft emission indicates the ejecta were
turning optically thin and therefore exposed to the hard radiation field of
the central white dwarf.  The consequent input of ionizing photons spreading
through the ejecta, counter-balanced for a short time the recombination
and delayed by $\sim$1 day the resumption of the fast decline, which is
driven by the dilution 
\clearpage

\IBVSfig{16cm}{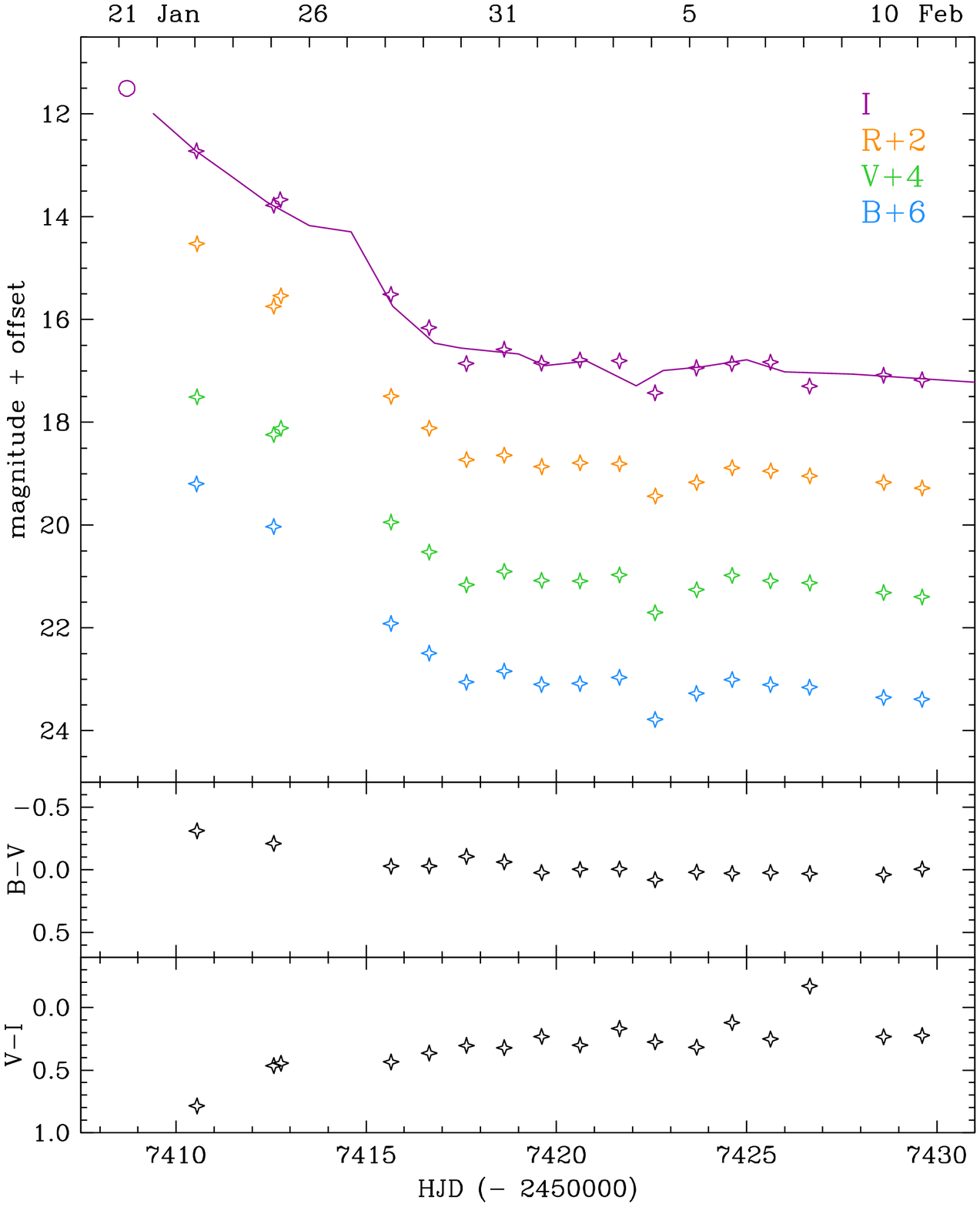}{Deconvolved SMARTS $B$$V$$R_{\rm C}$$I_{\rm C}$ 
photometry of the 2016 outburst of Nova LMC 1968. The open circle and the
line have the same meaning as in Figure~1.}

\noindent
of the rapidly expanding ejecta.  The fast decline lasted for only a few
more days, until the system brightness became sustained primarily by the
direct emission from the central white dwarf.  This phase corresponds to the
marked flattening of the light-curve after January 30.  As illustrated in
Figure~2, the brightness of the nova during this phase is identical in the
2016 and 2010 events, signifying identical conditions for the white dwarf
and the ongoing nuclear burning at its surface.  Intriguing is the dent
visible in all bands between February 3 and 4, similarly present in the 2010
light-curve.

For the 2010 eruption, Mroz et al.  (2014) report the nova being at normal
$I_{\rm C}$$\sim$19 mag quiescent brightness on 19.2 November and at $I_{\rm
C}$$\sim$12 mag (possibly saturated) on the next observation two days later,
on 21.2 November, and estimate in $I_{\rm C}$$\sim$11.5 the peak magnitude
supposedly reached around 20.2 November.  Mroz and Udalski (2016) quote
their $I_{\rm C}$=11.5 mag discovery observation on 21.2 January 2016 as
saturated.  Its position on the lightcurve of Figure~1 suggests however that
the saturation is probaby only marginal.  It is also at the same level of
the peak brightness estimated by Mroz et al.  (2014) for the 2010 event. 
Considering the identical 2010 and 2016 lightcurves, it seems reasonable to
assume that the nova reached its maximum $I_{\rm C}$$\sim$11.5 brightness on
21.2 January 2016.  Adopting for LMC a 18.5 mag distance modulus and a
$E_{B-V}$=0.08 mag reddening, the corresponding absolute magnitude is
$M_I$=$-$7.15 mag.  By low order extrapolation to 21.2 January of the color
evolutions depicted in Figures~1 and 2, the peak brightness attained by the
nova in the other bands can be estimated as $R_{\rm C}$$\sim$11.3,
$B$$\sim$12.0 and $V$$\sim$12.2.  As already noted by Sekiguchi et al. 
(1990) for the 1990 outburst, these magnitudes are much fainter than
expected - for LMC distance and reddening - on the basis of
magnitude-at-maximum/rate-of-decline (MMRD) relations, like the most recent
one by Downes and Duerbeck (2000) that predicts an observed $V_{\rm
peak}$$\sim$9 mag.

Only small portions of the light-curve were mapped during previous
outbursts.  Given the identical $I_{\rm C}$ light-curve for 2010 and 2016
eruptions, we may assume that the light-curves in other bands are similar
from outburst to outburst, in particular their decline rates.  The time
intervals elapsed between the 2016, 2010, 2002, 1990, and 1968 events are
then 1888.4, 2961.2, 4621.7, and 7728.3 days, respectively (uncertainties
$\pm$0.6 days).  Considering that ($i$) the 2010 outburst was missed
altogether by the community, and it was recovered only by inspecting years
later the archived OGLE data (Mroz et al.  2014), ($ii$) OGLE-IV is in
operation only by a few years, and ($iii$) inter-season gaps in the OGLE
monitoring of LMC last for $\sim$100 days, we may conclude that several more
outbursts of Nova LMC 1968 have probably gone missed.  A recurrence period
of $\sim$955 days would decently fit both the OGLE inter-season gaps and the
observed intervals between previous outbursts.  If correct, this would place
Nova LMC 1968 among those with the shortest known recurrence period, between
M31N 2008-12a ($\sim$1 year) and M31N 1963-09c ($\sim$5 years; Shafter et
al.  2015).

\references 

  Darnley, M. J., Williams, S. C. 2016, ATel 8616

  Downes, R.~A., Duerbeck, H.~W., 2000, AJ, 120, 2007

  Henden, A. et al. 2012, JAVSO 40, 430

  Kaur, A., et al. 2016, ATel 8626

  Liller, W. 1990, IAUC 4964

  Mroz, P., et al. 2014, MNRAS 443, 784

  Mroz, P., Udalski, A., 2016, ATel 8578

  Munari, U. et al.  2014, AJ 148, 81

  Page, K. L. et al. 2016, ATel 8615

  Pojmanski, G. 2002, ActA, 52,397

  Sekiguchi, K., et al. 1990, MNRAS 245, 28P

  Shafter, A. W. et al. 2015, ApJS 216, 34

  Shore, S. N. et al. 1991, ApJ 370, 193

  Sievers, J. 1970, IBVS 448

  Walter, F.~M., et al., 2012, PASP, 124, 1057 

\endreferences

\end{document}